# Dye stabilization and wavelength tunability in lasing fibers based on DNA

*Luana Persano[1], Adam Szukalski[2], Michele Gaio[3], Maria Moffa[1], Giacomo Salvadori[4], Lech Sznitko[2], Andrea Camposeo[1], Jaroslaw Mysliwiec[2], Riccardo Sapienza[3], Benedetta Mennucci[4], Dario Pisignano[1,5]*

[1]NEST, Istituto Nanoscienze-CNR and Scuola Normale Superiore, Piazza S. Silvestro 12, I-56127 Pisa, Italy.

[2]Faculty of Chemistry, Wroclaw University of Science and Technology, Wybrzeze Wyspianskiego 27, 50-370 Wroclaw, Poland.

[3]The Blackett Laboratory, Department of Physics, Imperial College London, London SW7 2AZ, UK.

[4]Department of Chemistry, University of Pisa, Via G. Moruzzi 13, I-56124 Pisa, Italy.

[5]Dipartimento di Fisica, Università di Pisa, Largo B. Pontecorvo 3, I-56127 Pisa, Italy.

**Abstract**

Lasers based on biological materials are attracting an increasing interest in view of their use in integrated and transient photonics. DNA as optical biopolymer in combination with highly-emissive dyes has been reported to have excellent potential in this respect, however achieving miniaturized lasing systems based on solid-state DNA shaped in different geometries to confine and enhance emission is still a challenge, and physico-chemical mechanisms originating fluorescence enhancement are not fully understood. Herein, a class of wavelength-tunable lasers based on DNA nanofibers is demonstrated, for which optical properties are highly controlled through the system morphology. A synergistic effect is highlighted at the basis of lasing action. Through a quantum chemical investigation, we show that the interaction of DNA with the encapsulated dye leads to hindered twisting and suppressed channels for the non-radiative decay. This is combined with effective waveguiding, optical gain, and tailored mode confinement to promote morphologically-controlled lasing in DNA-based nanofibers. The results establish design rules for the development of bright and tunable nanolasers and optical networks based on DNA nanostructures.

**1. Introduction**

Supramolecular chemistry and nanotechnology approaches based on deoxyribonucleic acid (DNA)[1] are prominent examples of how biological macromolecules can be used to assemble complex structures. This concept has been gaining a continuously increasing interest for the generation of multicomponent systems for photonics and optoelectronics.[2] Applications demonstrated to date include using DNA as electron blocking layer in organic light-emitting devices,[3] and as template for non-linear optical materials.[4] Furthermore, the use of DNA in organic micro- and nano-lasers[5] is especially appealing, since embedded chromophores might show enhanced performance compared to blends with other synthetic polymers.[2d,5b,6] Such systems, primarily if equipped with effective chemo-structural tools for spectral control,[7] can lead to the development of optical communication chips and high-throughput chemical and biological sensors. The encapsulation in confining structures or matrices suitable to conveniently tailor the energy levels of lasing dyes can be highly relevant in this respect, as demonstrated in shape-engineered microlaser based on metal-organic frameworks.[7c, 8] In addition, DNA is available as a waste product of the fishing industry, thereby potentially embodying a sustainable building block for realizing transient photonic components.[5d]

While a few studies reported that dyes such as sulforhodamine[5b] and hemicyanine[2d,6] show improved fluorescence in DNA, mechanisms at the base of these properties remain elusive due to the concomitant effects from intercalation or groove-binding as well as from microstructure cavity morphologies. Among other solid-state structures, DNA-based micro- and nanofibers might offer significant advantages in this respect, including highly effective control of sustained optical modes in the fibrous geometry. Light scattering and confinement in the complex microstructure of the biopolymer filaments, together with potential dye stabilization resulting in the significant reduction of non-radiative decay channels for excitations, would make light-emitting DNA-based fibers ideal systems for lasing action. So far, dye-doped fibers of DNA complexed with the cationic surfactant cetyltrimethylammonium chloride (CTMA) have been reported as fluorescent material,[6] as white emitters upon donor-acceptor resonance energy transfer,[2c] and as all-optical switches through incorporation of a push–pull, π-conjugated pyrazoline derivative featuring intramolecular charge transfer,[4b] but laser systems based on such class of nanofibers are still unexplored.

In this work, we clarify the dual aspects of fluorescence enhancement due to DNA interaction for the lasing dye, [3-(2,2-dicyanoethenyl)-1-phenyl-4,5-dihydro-1H-pyrazole] (DCNP), and the behavior of optical modes due to confinement in the assembled fibrous architecture. Time-Dependent Density Functional Theory (TD-DFT) calculations are also performed for DCNP in gas-phase, in a polar solvent and when intercalated in a DNA fragment. The found synergistic effects lead to measure for the first time lasing action from the DNA-based fibers, with remarkable wavelength tunability and emission regime controllably switchable from multimodal to broadband lasing.

## 2. Results and Discussion

Arrays of uniaxially-aligned DCNP/DNA-CTMA fibers with ~1% dye loading in weight and sample size of many mm$^2$, spun from ethanol:chloroform (3:1 v/v) solutions with biopolymer concentration in the range 2-4% (w/w) are shown in Figure 1a,b (see the Experimental Section for details). The molecular structure of DCNP is shown in Figure S1 in the Supporting Information. Individual filaments often show a bi-modal distribution of their transversal size, with components peaked in the scale of 1 to a few μm and at few hundreds of nm, respectively (Figure 1c), and are mainly ribbon-shaped as highlighted by atomic force microscopy (AFM, Figure 1d-g). The average value of measured transversal size ranges between 400 nm and 1.7 μm (Figure S2) upon increasing the DNA concentration in the original solution, which leads to markedly different lasing properties as explained below. The absorption, photoluminescence, and amplified spontaneous emission (ASE) spectra of DCNP in the DNA-CTMA solid-state matrix are shown in Figure S3. Fibers exhibit bright and uniform fluorescence, which suggests that stable amounts of dyes are present along their length (Figure 1h).

Experimental studies and previous quantum mechanical calculations have shown that DCNP presents a photophysics which depends on the polarity and viscosity of the environment.[9] A brighter fluorescence is found for nonpolar solvents and in more viscous environments, and ASE has been observed in a poly(methyl methacrylate) matrix,[10] suggesting DCNP as suitable dye for use in rigid matrices. In fact, DCNP presents the typical characteristics of a potential twisted-internal-charge-transfer (TICT) system: an electron donor group (benzene ring), an electron attractor group (cyan groups) and a network of conjugated bonds that delocalize the electronic density, connecting the two sub-units. The photophysics of TICT systems is determined by important geometrical changes in the excited state that involve a rotation between donor and acceptor groups. When the absorption of a photon occurs, a vertical excited state is reached known as 'LE' (locally excited). If the donor-acceptor units are free to rotate, the LE state relaxes into the excited TICT state from which the nonradiative decay will be possibly activated.

There are two stable isomers of DCNP, *S-cis* and *S-trans*, obtained following a rotation around the dihedral NC1-C2C3 (Figure S1). From the calculations, we obtain that both isomers have a planar geometry in the ground state, with the *S-cis* isomer being less stable than the *S-trans* isomer, but the difference in free energy is reduced when the effects of a polar solvent (acetonitrile, ACN) are included using the Polarizable Continuum Model (PCM)[11](Table 1). The isomeric forms of DCNP are separated from each other by a sizeable barrier of about 0.6 eV for the ground state. This barrier decreases only slightly in the lowest singlet excited state $S_1$ but the energetic stabilization of the *cis*/*trans* forms is reversed with the *S-cis* becoming the most stable isomer. In addition, both isomers exhibit strong dipole allowed transitions to $S_1$, separated from the optically forbidden (dark) $S_2$ state. Fluorescence from the $S_1$ state of both isomers is fully allowed and a remarkable red-shift of the fluorescence for the *S-cis* isomer with respect to the *S-trans* one is predicted (Table 1).

To investigate the possible TICT behavior, relaxed scans are conducted around the C1C2=C3C4 dihedral angle in the $S_1$ excited state for the isolated molecule and the solvated one in ACN. Results support the suggestion that the occurrence of the deactivation process can be linked to the excited-state flexibility of the molecule along the CC=CC torsion. A discontinuity is found in the energy of the lowest excited state at about 60° of rotation (Figure 2a) which is accompanied by an abrupt change in the nature of the state: for a torsion angle below 60° $S_1$ is a bright state, while after 60° it becomes dark as confirmed by the corresponding oscillator strengths. The energy barrier to get to the discontinuity decreases upon moving from the isolated (0.379 eV) to the solvated molecule (0.353 eV for the *trans* and 0.280 for the *cis* isomer). Along this CC=CC torsion the energy of the ground state rises and a conical intersection between the ground and the lowest excited state is expected near the perpendicular orientation. It is, however, important to point out that conical intersections cannot be properly described by the TDDFT approach here used.

For the $S_1$ state, increasing the degree of torsion of the dihedral CC=CC breaks the delocalization, and at 90° one no longer has a delocalized system, but a system in which the negative charge is concentrated on the cyan groups. This change in the nature of the state along the torsion is here quantified in terms of the Natural Transition Orbitals (NTO)[12] mainly involved in the $S_0$-$S_1$ transition (Figure 2b). These data are in agreement with the hypothesis that rotation around the dihedral CC=CC can origin fluorescence quenching. Therefore, in DNA-induced conditions that hinder the twisting, a higher fluorescence yield is expected, as this non-radiative decay channel will be inhibited.

To investigate this hypothesis, we extend the calculations to DCNP intercalated in a model DNA fragment. For this analysis, only *S-trans* conformation of DCNP is considered. Two different configurations are selected after a preliminary search on twenty different intercalation modes (see Experimental for details), *i.e.* sandwich-like (Figure 3a) and rotated (Figure 3b). The structure of DCNP in the two configurations is optimized for both the ground and the lowest excited state, and Table 2 reports the values of the dihedral angle, the relative free energy and the absorption and fluorescence energies. Due to the simplified model used for the DNA, where the fully atomistic description is limited the two pairs of adjacent bases, the calculated free energies should be used only to compare the relative stability of the two intercalation modes. In fact, it is expected that their differences are mainly due to stacking and electrostatic interactions, both of which are properly accounted for in our model combining DFT with empirical corrections for dispersion effects (see Experimental for details). From the results reported in Table 2, we see that for the ground state, the rotated configuration is significantly higher in energy with respect to the sandwich configuration: the greater stability of the latter is explained in terms of more effective stacking interactions and more favorable electrostatic interactions, since the DNCP benzene group, which has a partially positive charge, is close to a negatively charged phosphate group.

Due to this larger stability, we focus only the sandwich configuration, finding that only a slight shift in absorption and fluorescence energies occurs due to the intercalation with respect to the solvated system. What changes dramatically moving from the solvated to the intercalated system is the energy associated with the torsion around the dihedral CC=CC. By repeating the scan of the $S_1$ energy along such dihedral angle, we in fact see that, due to the steric hindrance, geometry optimizations could not be successfully terminated with angles larger than 50°. However, it is still possible to make some comparisons with the molecule in ACN. In the case of intercalation, we observe that reaching a 50° torsion of the dihedral angle leads to an energy increase of 0.424 eV, whereas in the case of DCNP in ACN the increase in energy is 0.252 eV to reach the same degree of torsion. Such a steeper increase of the energy for the torsion in DNA clearly indicates that the non-radiative decay channel, suggested in the non-intercalated system, cannot be effective when DNA is present. This is in agreement with the hypothesis that a constrained environment can enhance the fluorescence performance of the light-emitting system. Therefore, micro- and nanostructures based on DNA and embedding the DCNP molecule are excellent candidates to exhibit bright fluorescence and lasing properties.

In addition, the optical properties of DCNP/DNA-CTMA fibrous assemblies depend on both individual filaments and of their collective morphology. The critical thickness for disordered lasing from organic nanostructured slabs is $t_{CR} = \pi(l_G \, l_T/3)^{1/2}$, where $l_G$ and $l_T$ are the gain length and the transport mean free path of the diffused photons, respectively.[13a] Typical values for lasing dyes incorporated in organic matrices, i.e. $l_G$ of few hundreds of μm and $l_T$ in the range 2-10 μm, leads to $t_{CR}$ of about 100 μm,[13b] which is straightforwardly achieved in DCNP/DNA-CTMA assemblies. At individual fiber level, the different morphologies of single DCNP/DNA-CTMA fibers would lead to varied lasing emission characteristics. This aspect is strongly related to the waveguiding features of DNA-based fibers, that are analyzed by microphotoluminescence (Figure S4). The effective waveguiding of the light emitted by DCNP and channeled along single DNA-CTMA fibers can be assessed by recording the photoluminescence intensity ($I_{PL}$) from the fiber body as a function of the distance ($d$) of the probed segment from the excitation area. Fitting the experimental data by an exponential function, $I_{PL}=I_0\times\exp(-\alpha d)$, where $I_0$ is the intensity measured at very small $d$ values, and $\alpha$ is the loss coefficient, leads to $\alpha= 100$ cm$^{-1}$, corresponding to a light transport length of about 100 μm. These values are lower than the losses measured in nanofibers doped with CdSe quantum dots, and comparable with the best values reported for fibers clearly embedding MoS$_2$ or made of synthetic, conjugated polymers.[14] Effective waveguiding along the longitudinal axis of the DCNP/DNA-CTMA fibers might be especially important for promoting optical gain along the array length, which can in turn lead to lasing. Indeed, upon excitation with about 10 ns pulses at 355 nm and with a stripe pumping geometry, DCNP/DNA-CTMA arrays of fibers with average transversal size of 400 nm show multimodal lasing, with several spectrally-decoupled and narrow (full width at half maximum, FWHM $\cong$ 0.2 nm) emission peaks (Figure 4a). Upon increasing the average value of the fiber transversal size, this behavior continuously changes to a broadband lasing regime, featuring a single, smooth emission band that is only slightly featured at its top (Figure 4a). This can be explained with an increase of the spectral density of the lasing modes coalescing into a single broad peak when their spacing is less than their width.

It is noteworthy that the critical parameter determining the lasing regime is the confinement factor,[15] $\eta$, supported in the DCNP/DNA-CTMA fibers (Figure 4b). $\eta$ indicates the fraction of the field intensity that is confined within the fiber, namely the amount of power emitted by the DCNP dye that is retained within the DNA-CTMA fiber. For fibers with small transversal size, it is expected that the fundamental mode extends out of the fiber body, thus leading to a decrease of the value of $\eta$.[16,17] Indeed, in the simplified case of a cylindrical fiber with diameter $d_F$, $\eta$ is calculated by the expression of the fractional guided power in the fundamental mode,[15] $\eta = 1-[(2.4e^{-1/V})^2/V^3]$, with $V = (\pi d_F/\lambda_p)\times(n_F^2-1)^{1/2}$ where $\lambda_p$ is the emission wavelength and $n_F$ is the refractive index of DNA ($\cong$1.5). A transition from multimode to broad, single-band lasing is found for the DCNP/DNA-CTMA fibers upon increasing the fiber diameter above $\lambda_p$, which corresponds to an increase of the confinement factor that closely approaches unity. Overall, a remarkable wavelength tunability is achieved, namely a lasing wavelength varying from 630 nm to about 655 nm depending on the fiber morphology. The observed red-shift of the lasing wavelength upon increasing the transversal size of the fibers is related to the degree of overlap of the guided modes with the gain medium.

For larger fibers, the guided mode is effectively contained in the fiber ($\eta$>99%), which occurs for all the wavelengths in the spectral interval of the DCNP optical gain (620-660 nm). This condition clearly leads to lasing closely to the wavelength of maximum optical gain. Upon decreasing the fiber size, the fraction of the mode that is effectively overlapped with the gain medium is decreased as well, and this effect is more critical for longer wavelengths. For instance, for 400 nm fibers, $\eta$ decreases from 0.80 to 0.77 upon increasing the wavelength in the range of the DCNP optical gain. Therefore, lasing is correspondingly favored at shorter wavelengths. The relative intensity of individual narrow peaks can change due to mode competition for lasing (Figure 5a,b). Excitation thresholds are found for the lasing emission, at about 11 mJ cm$^{-2}$ and in the range 30-40 mJ cm$^{-2}$ for DCNP/DNA-CTMA arrays of fibers with average diameter 1.7 µm and 400 nm, respectively (Figure 5c,d).

To rationalize the special lasing behaviour of the DCNP/DNA-CTMA ribbon-shaped nanofibers, we also calculate numerically the profile of sustained optical modes. In fact, the modes of an individual fiber can be calculated analytically for a cylindrical geometry,[15] whereas our DNA-based fibers generally show an elliptical cross-section, where the two dimensions $r_1$ and $r_2$ in the inset of Figure 6a are different. To analyze this system, we use a finite-difference time-domain solver (Lumerical) and evaluate the resonant modes for frequencies close to the DCNP optical gain spectrum, at 640 nm, for typical ellipticity, $r_2/r_1$ = 0.5, and a refractive index, $n_F$. Figure 6a highlights the dispersion of the first ten modes as a function of the transverse momentum, $k$, for varied fiber size, normalized to the light wavelength. Each mode is labelled with a different color and is found to appear for large enough fibers (i.e. for high enough $r_2$ values). The guided mode dispersion lies in between the light line in air (left dashed line) and in the polymer (right dashed line), i.e. they cannot propagate in air as their momentum exceeds the maximum in air. The transverse intensity profiles of the first ten modes is displayed in Figure 6b. It is worthwhile pointing out that the ellipticity of the fibers splits the degeneracy between the optical modes polarized along the two perpendicular directions in the fiber section, which is why the modes structure in Figure 6a is composed of pairs of closely spaced modes. This is also well seen in the spatial profile of the modes. The guided modes are all well-confined in the DCNP/DNA-CTMA fibers, thus overlapping with the DCNP gain medium. As characteristic of lasing fiber architectures, the fundamental mode, which has a Gaussian shape, namely a better confinement and overlap with the gain medium, is the most likely to lase first. The modal gain, calculated for the different modes and normalized for the gain of a plane wave at the gain peak wavelength (645 nm), is plotted in Figure 6c. This shows how the gain grows for larger fibers, i.e. for a large amount of DCNP available to each mode. In this respect, the spectral features found for lasing DCNP/DNA-CTMA arrays appear to be mainly affected by single-fiber features and particularly modal gain. This is highly appealing for potential applications in sensing optical components, since both the fiber refractive index and the shape have a profound impact on the modal gain and might be effectively changed by molecular diffusion or swelling of the DNA matrix.

## 3. Conclusion

In summary, we have reported that lasers with remarkable wavelength tunability, and with regime varying between multimodal and broadband action can be realized based on DNA fibers incorporating a DCNP dye. With the help of quantum chemical calculations, we have shown that the DCNP interaction with DNA effectively hinders the excited state geometrical distortions responsible to non-radiative decay channels thus leading to enhanced light emission. The finding, that upon suppressed twisting the non-radiative decay is disfavoured as well, has general validity once an enough viscous environment or steric hindrance are provided in the host micro- or nanostructures. This makes DCNP/DNA-CTMA fibers particularly promising for enhancing light emission as both constraining conditions are in place for the dye. On top of this, the effective waveguiding, optical gain, and tailored mode confinement in DNA nanofibers work as supporting mechanisms to promote lasing action with morphology-controlled properties. These results represent an important step toward the design of miniaturized, highly-controlled lasers based on DNA nanostructures, as well as toward the rationalization of the various effects promoting enhanced emission in the same.

## 4. Experimental Section

*DCNP/DNA-CTMA fibers*. DNA (about 2,000 bp, Sigma-Aldrich) functionalized with the CTMA surfactant (Sigma-Aldrich) and doped with DCNP (molecular structure in Figure S1) was used. For realizing DCNP/DNA-CTMA fibers, the DNA-CTMA complex was dissolved in a mixture of ethanol and chloroform (3:1 v/v), with a polymer to solvent weight ratio varying in the range 2-4% (w/w), and the obtained solution was stirred for 24 h. A 1% concentration of DCNP was used in the dry mass of biopolymer. Solutions were electrospun using a syringe with stainless steel needle and feeding rate of 0.5 mL/h (33 Dual Syringe Pump, Harvard Apparatus Inc.), a distance between the needle and collector surface of 20 cm, and an applied voltage of 10 kV. To generate aligned fibers, a rotating cylinder was used as collector. The morphology of the doped fibers was investigated by scanning electron microscopy (SEM, Nova NanoSEM 450, FEI) with an accelerating voltage of 3 kV and an aperture size of 30 μm, following thermal deposition of 5 nm of Cr (PVD75, Kurt J. Lesker Co.). The average diameter of the fibers was calculated from the SEM micrographs using an imaging software. Atomic force microscopy on individual DCNP/DNA-CTMA fibers was performed by a Multimode system (Veeco) equipped with a Nanoscope IIIa controller and working in tapping mode.

*Confocal microscopy, single-fiber waveguiding experiments, and spectroscopy.* Confocal fluorescence micrographs were obtained by exciting the sample with a diode laser ($\lambda$ = 405 nm), focused through an objective lens (20×, numerical aperture = 0.5). The emission from the fibers was then collected by the same objective and analyzed by a multianode photomultiplier. A spectrophotometer (Perkin Elmer, Lambda 950) was used for transmittance measurements. Lasing was achieved by pumping samples with the third harmonic ($\lambda$ = 355 nm) of a Nd:YAG laser beam (repetition rate = 10 Hz, pulse duration $\cong$ 10 ns), focused in a rectangular stripe (400 µm×1 cm). The emission was then collected from the sample edge, coupled to an optical fiber and analyzed by a monochromator (iHR320, Jobin Yvon) equipped with a CCD detector (Symphony, Jobin Yvon). Lasing thresholds were measured by systematically varying the excitation fluence. The waveguiding properties of individual DCNP/DNA-CTMA fibers were analyzed by using a micro-photoluminescence (µ-PL) setup, based on an inverted microscope (IX71, Olympus) equipped with a 60× oil immersion objective (NA=1.42, Olympus) and a CCD camera. The PL was excited by the diode laser coupled to the microscope through a dichroic mirror and focused on the sample by the objective. Part of the light emitted by the DCNP chromophore in the DNA-based fiber, excited by the tightly focused laser spot, was coupled into the fiber and waveguided. The optical losses coefficient was then measured upon collecting images of the intensity of emission diffused by the fiber surface, and by analyzing the spatial decay of emission as a function of the distance from the exciting laser spot.[14a]

*Density Functional Theory*. Calculations were performed with DFT and its TD extension for excited states. For the isolated and the solvated DCNP, geometry optimizations in the ground state have been performed at B3LYP/6-31G(d) level. Solvent effects were included through the Integral Equation Formalism[11] version of the PCM (IEF-PCM).[18] All calculations of DCNP intercalated in DNA were performed using a simplified model that limits the interactions between DCNP and DNA to the two pairs of adjacent bases (GC-AT), the deoxyribose sugars and the phosphate groups that connect these bases. The hydrogens of the OH present in the sugar have been replaced with neutral groups -CH$_3$. The complex formed by DCNP and the DNA fragment has been embedded in a cavity of a dielectric medium to mimic the effect of the solvent. Different intercalation modes were generated for DCNP using the criterium of maximizing its stacking and electrostatic interactions with DNA for both sandwich-like and rotated configurations. These 20 initial structures were subjected to a geometry optimization with the semi-empirical PM6 method and finally compared to eliminate eventual replicas. The remaining 12 structures are shown in **Figure S5**. The two structures showing the lowest energy for the sandwich-like and for the rotated configuration (reported in the two boxes) were finally re-optimized at the same DFT description used for the solvated system (B3LYP/6-31G(d)) but with the addition of empirical dispersion corrections.[19] In the optimization of these intercalated systems, the geometry of the DNA pocket was kept frozen and described using a 6-31G basis set. For both solvated and intercalated systems, excited state calculations have been performed at TD-DFT level using the CAM-B3LYP[20] functional in combination with 6-31+G(d) basis set: this functional is in fact known to properly describe charge-transfer states. Relaxed excited state potential energy surfaces (PESs) have been obtained assuming a complete relaxation of the solvent polarization. To account for a proper response of the solvent, a State-Specific correction has been introduced for each point of the PESs through the corrected Linear Response formulation.[21] All calculations have been done with Gaussian '16 software.[22]

*Optical properties*. The optical modes supported in the fibers were calculated using a commercial mode solver as provided by Lumerical. This is a numerical solution of the Maxwell's equations in the fiber geometry, to find the eigenmodes of the structure. The fiber structures simulated were elliptical in their cross-section, with semi-minor and semi-major axis $r_1$ and $r_2$, respectively. The calculations were performed for a constant ellipticity $r_2/r_1$ = 0.5. Intensity profiles for typical modes were calculated for fibers with $r_1$=0.25 µm and $r_2$ = 0.5 µm. From the numerical simulations we obtained the electric field **E** in the fiber and its wavevector $k'$, $\mathbf{E}(x,y,z;\lambda) = \mathbf{E}(x,y,z_0; \lambda)*\exp(-ik'z)$, from which we estimate the gain as $[\text{Im}(k')/\text{Im}(k)]^2$, where $k$ is the wavevector in the bulk biopolymer.

**Acknowledgements**

The research leading to these results has received funding from the European Research Council under the European Union's Horizon 2020 Research and Innovation Programme (Grant Agreement n. 682157, "*x*PRINT"), and from the Italian Minister of University and Research PRIN 2017PHRM8X and PRIN 201795SBA3 projects. D.P. acknowledges the support from the project PRA_2018_34 ("ANISE") from the University of Pisa. A.S. acknowledges the support from the Foundation for Polish Science (FNP). J.M. acknowledges the support of The National Science Center, Poland (2016/21/B/ST8/00468) and of the Wroclaw University of Science and Technology. R.S. acknowledges support from The Leverhulme Trust (RPG-2014-238) and the Royal Society (IE160502). B.M. acknowledges funding from the European Research Council under the grant ERC-AdG-786714 (LIFETimeS).

**References**

[1] a) *DNA in Supramolecular Chemistry and Nanotechnology* (Eds.: E. Stulz, G. H. Clever), John Wiley & Sons, Ltd, Chichester, **2015**; b) N. C. Seeman, *Nano Lett.* **2020**, *20*, 1477-1478.

[2] a) A. J. Steckl, *Nat. Photonics* **2007**, *1*, 3-5; b) Y.-W. Kwon, C. H. Lee, D.-H. Choi, J.-I. Jin, *J. Mater. Chem.* **2009**, *19*, 1353-1380; c) Y. Ner, J. G. Grote, J. A. Stuart, G. A. Sotzing, *Angew. Chem. Int. Ed.* **2009**, *48*, 5134-5138; d) Y. Ner, D. Navarathne, D. M. Niedzwiedzki, J. G. Grote, A. V. Dobrynin, H. A. Frank, G. A. Sotzing, *Appl. Phys. Lett.* **2009**, *95*, 263701; e) S. H. Back, J. H. Park, C. Cui, D. J. Ahn, *Nat. Commun.* **2015**, *7*, 20134.

[3] J. A. Hagen, W. Li, A. J. Steckl, *Appl. Phys. Lett.* **2006**, *88*, 171109.

[4] a) D. Wanapun, V. J. Hall, N. J. Begue, J. G. Grote, G. J. Simpson, *ChemPhysChem* **2009**, *10*, 2674-2678; b) A. Szukalski, M. Moffa, A. Camposeo, D. Pisignano, J. Mysliwiec, *J. Mater. Chem. C* **2019**, 7, 170-176.

[5] a) Y. Kawabe, L. Wang, T. Nakamura, N. Ogata, *Appl. Phys. Lett.* **2002**, *81*, 1372-1374; b) Z. Yu, W. Li, J. A. Hagen, Y. Zhou, D. Klotzkin, J. G. Grote, A. J. Steckl, *Appl. Opt.* **2007**, *46*, 1507-1513; c) M. Leonetti, R. Sapienza, M. Ibisate, C. Conti, C. López, *Opt. Lett.* **2009**, *34*, 3764-3766; d) A. Camposeo, P. Del Carro, L. Persano, K. Cyprych, A. Szukalski, L. Sznitko, J. Mysliwiec, D. Pisignano, *ACS Nano* **2014**, *8*, 10893-10898.

[6] Y. Ner, J. G. Grote, J. A. Stuart, G. A. Sotzing, *Soft Matter* **2008**, *4*, 1448-1453.

[7] a) W. Zhang, Y. Yan, J. Gu, J. Yao, Y. S. Zhao, *Angew. Chem. Int. Ed.* **2015**, *54*, 7125-7129; b) A. J. C. Kuehne, M. C. Gather, *Chem. Rev.* **2016**, *116*, 12823-12864; c) Y. Lv, Z. Xiong, H. Dong, C. Wei, Y. Yang, A. Ren, Z. Yao, Y. Li, S. Xiang, Z. Zhang, Y. S. Zhao, *Nano Lett.* **2020**, *20*, 2020-2025.

[8]   a) Y. Wei, H. Dong, C. Wei, W. Zhang, Y. Yan, Y. S. Zhao, *Adv. Mater*. **2016**, *28*, 7424–7429. b) H. He, H. Li, Y. Cui, G. Qian, *Adv. Opt. Mater*. **2019**, *7*, 1900077.

[9]   O. Morawski, B. Kozankiewicz, A. Miniewicz, A. L. Sobolewski. *ChemPhysChem* **2015**, *16*, 3500-3510.

[10]  L. Sznitko, J. Mysliwiec, K. Parafiniuk, A. Szukalski, K. Palewska, S. Bartkiewicz, A. Miniewicz, *Chem. Phys. Lett*. **2011**, *512*, 247–250.

[11]  E. Cancès, B. Mennucci, J. Tomasi, *J. Chem. Phys*. **1997**, *107*, 3032–3041.

[12]  R. L. Martin, *J. Chem. Phys* **2003**, *118*, 4775-4777.

[13]  a) D. S. Wiersma, A. Lagendijk, *Phys. Rev. E* **1996**, *54*, 4256-4265; b) M. Montinaro, V. Resta, A. Camposeo, M. Moffa, G. Morello, L. Persano, K. Kazlauskas, S. Jursenas, A. Tomkeviciene, J. V. Grazulevicius, D. Pisignano, *ACS Photonics* **2018**, *5*, 1026-1033.

[14]  a) V. Fasano, A. Polini, G. Morello, M. Moffa, A. Camposeo, D. Pisignano, *Macromolecules* **2013**, *46*, 5935-5942; b) A. Portone, L. Romano, V. Fasano, R. Di Corato, A. Camposeo, F. Fabbri, F. Cardarelli, D. Pisignano, L. Persano, *Nanoscale* **2018**, *10*, 21748-21754.

[15]  A.W. Snyder, J. Love, *Optical Waveguide Theory*, Chapman and Hall, London, **1983**.

[16]  J. C. Johnson, H. Yan, P. Yang, R. J. Saykally, *J. Phys. Chem. B* **2003**, *107*, 8816.

[17]  D. O'Carroll, I. Lieberwirth, G. Redmond, *Nat. Nanotechnol.* **2007**, *2*, 180.

[18]  J. Tomasi, B. Mennucci, R. Cammi, *Chem. Rev.* **2005,** *105,* 2999–3093.

[19]  S. Grimme, S. J. Antony, S. Ehrlich, H. Krieg, *J. Chem. Phys*. **2010**, *132*, 154104.

[20]  T. Yanai, D.P. Tew, N.C. Handy*, Chem. Phys. Lett*. **2004**, *393*, 51–57.

[21]  M. Caricato, B. Mennucci, J. Tomasi, F. Ingrosso, R. Cammi, S. Corni, G. Scalmani, *J. Chem. Phys*. **2006**, *124*, 124520–124513.

[22]  *Gaussian 16, Revision C.01*, M. J. Frisch, G. W. Trucks, H. B. Schlegel, G. E. Scuseria, M. A. Robb, J. R. Cheeseman, G. Scalmani, V. Barone, G. A. Petersson, H. Nakatsuji,

X. Li, M. Caricato, A. V. Marenich, J. Bloino, B. G. Janesko, R. Gomperts, B. Mennucci, H. P. Hratchian, J. V. Ortiz, A. F. Izmaylov, J. L. Sonnenberg, D. Williams-Young, F. Ding, F. Lipparini, F. Egidi, J. Goings, B. Peng, A. Petrone, T. Henderson, D. Ranasinghe, V. G. Zakrzewski, J. Gao, N. Rega, G. Zheng, W. Liang, M. Hada, M. Ehara, K. Toyota, R. Fukuda, J. Hasegawa, M. Ishida, T. Nakajima, Y. Honda, O. Kitao, H. Nakai, T. Vreven, K. Throssell, J. A. Montgomery, Jr., J. E. Peralta, F. Ogliaro, M. J. Bearpark, J. J. Heyd, E. N. Brothers, K. N. Kudin, V. N. Staroverov, T. A. Keith, R. Kobayashi, J. Normand, K. Raghavachari, A. P. Rendell, J. C. Burant, S. S. Iyengar, J. Tomasi, M. Cossi, J. M. Millam, M. Klene, C. Adamo, R. Cammi, J. W. Ochterski, R. L. Martin, K. Morokuma, O. Farkas, J. B. Foresman, D. J. Fox, Gaussian, Inc., Wallingford CT, **2016**.

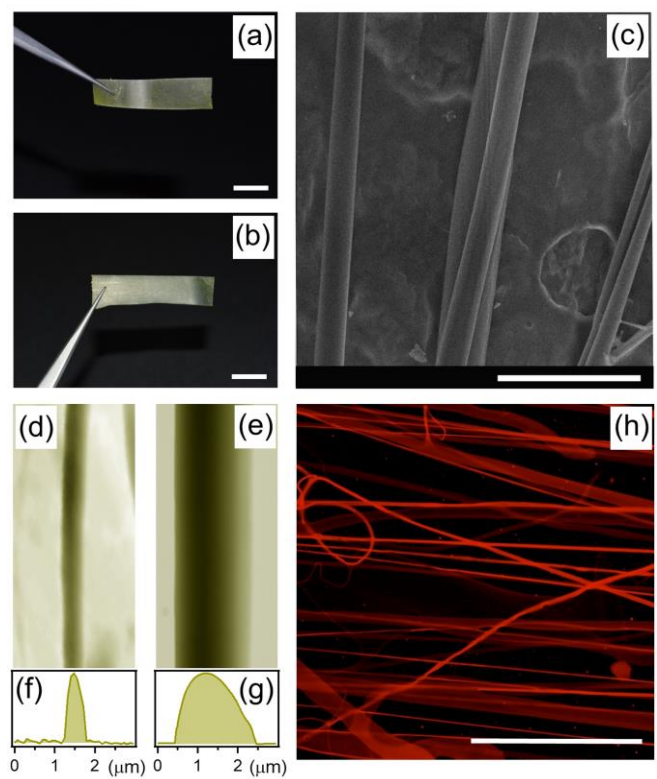

**Figure 1.** Exemplary photographs (a,b), SEM micrograph (c), AFM topographic images (d,e) and corresponding height profiles (f,g), and confocal fluorescence micrograph (h) of DCNP/DNA-CTMA fibers. Scale bars: 5 mm (a,b); 10 µm (c); 300 µm (h). Imaged samples are realized by a DNA-CTMA to solvent ratio from 2% (a,d,f) to 4% in weight (b,c,e,g,h). The fiber height shown in AFM micrograph in (f) and (g) is 200 nm and 1 µm, respectively.

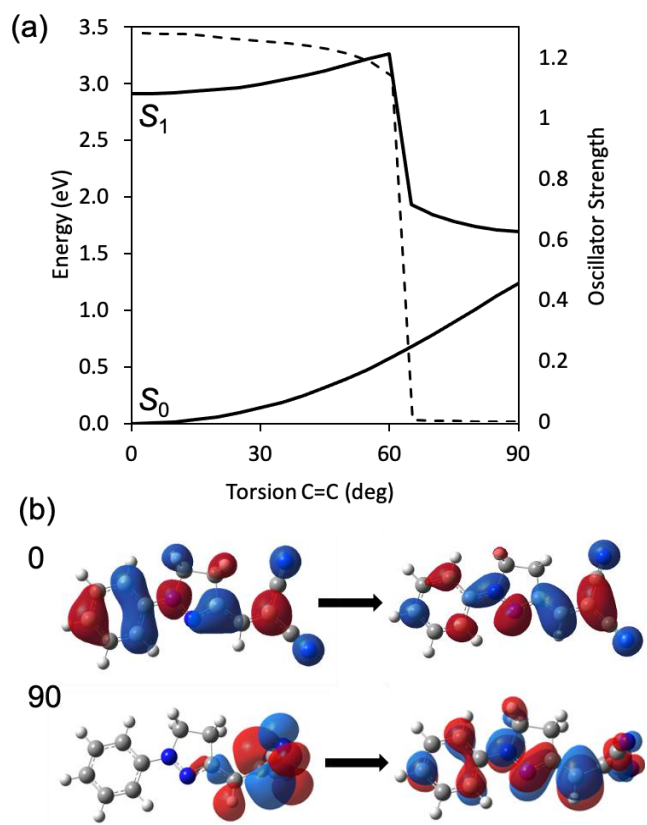

**Figure 2.** (a) Energy of the lowest excited ($S_1$) and the ground ($S_0$) states of DCNP for the *S-trans* isomer in ACN, along the rotation of the C1C2=C3C4 dihedral angle. Oscillator strength along the rotation is also reported (dashed line). (b) Main NTO for the $S_0 \rightarrow S_1$ transition at planar (0) and twisted (90) structure.

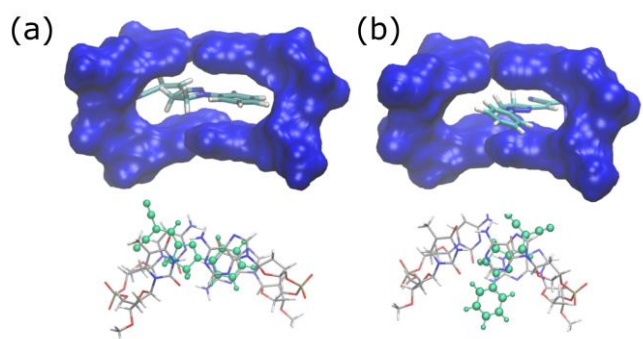

**Figure 3.** Two different representations of the two modes of intercalation of DCNP in DNA. (a) Sandwich-like configuration. (b) Rotated configuration. In the upper panels we report the surface representing the binding pocket created by the two base pairs and the deoxyribose sugars and the phosphate groups that connect them, reported in the lower panels.

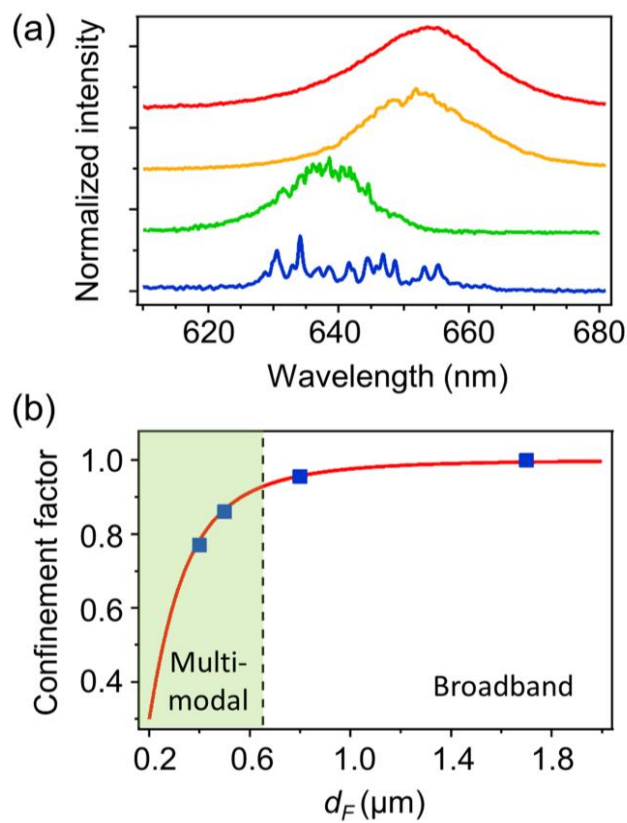

**Figure 4.** (a) Different emission spectra of DCNP/DNA-CTMA fibers with average transversal sizes 400 nm, 500 nm, 800 nm, and 1.7 μm (from bottom to top, respectively). (b) Corresponding diagram of lasing regimes. Red line: confinement factor *vs.* fiber diameter. Blue dots: $\eta$ values calculated for fibers with transversal size corresponding to average values measured for DCNP/DNA-CTMA. The vertical dashed line marks the $d_F$ value equal to a 650 nm emission peak wavelength. Multimodal vs. broadband transition occurs when the transversal size of DCNP/DNA-CTMA fibers equals the emission wavelength, $\lambda_p$ ($\eta \geq 0.9$).

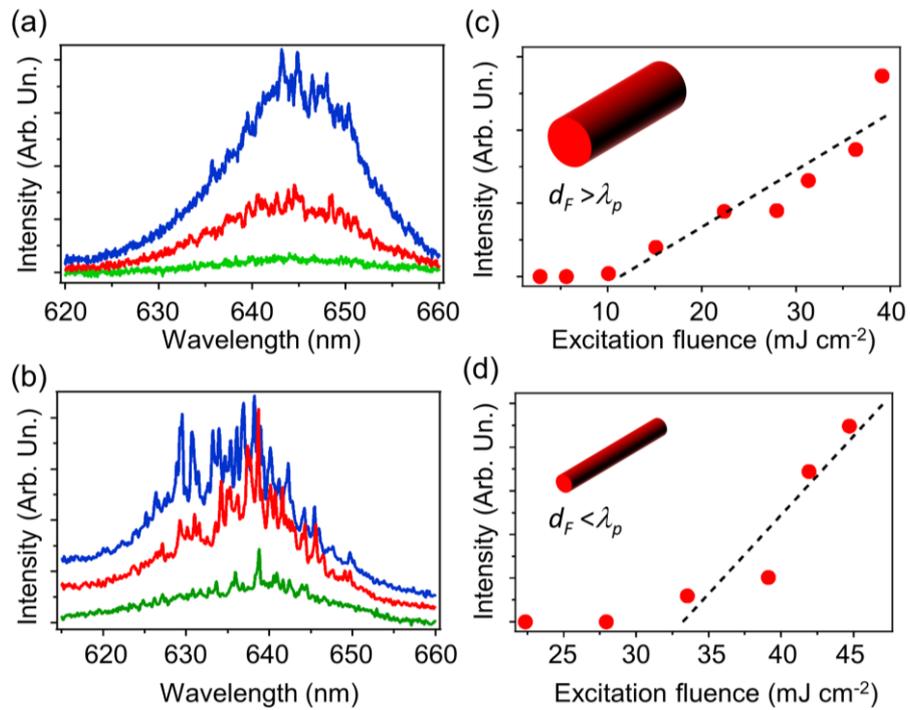

**Figure 5.** (a,b) Single-shot spectra detected from DCNP/DNA-CTMA fibers of different average transversal size, measured upon varying the excitation fluence. $d_F$ = 1.7 µm (a); 400 nm (b). Excitation fluence values, from bottom to top: (a) 10, 15 and 30 mJ cm$^{-2}$; (b) 30, 40, 45 mJ cm$^{-2}$. (c,d) Corresponding lasing emission intensity *vs*. excitation fluence. Insets: schematics of the two classes of DCNP/DNA-CTMA fibers, with $d_F > \lambda_p$ and $d_F < \lambda_p$, respectively.

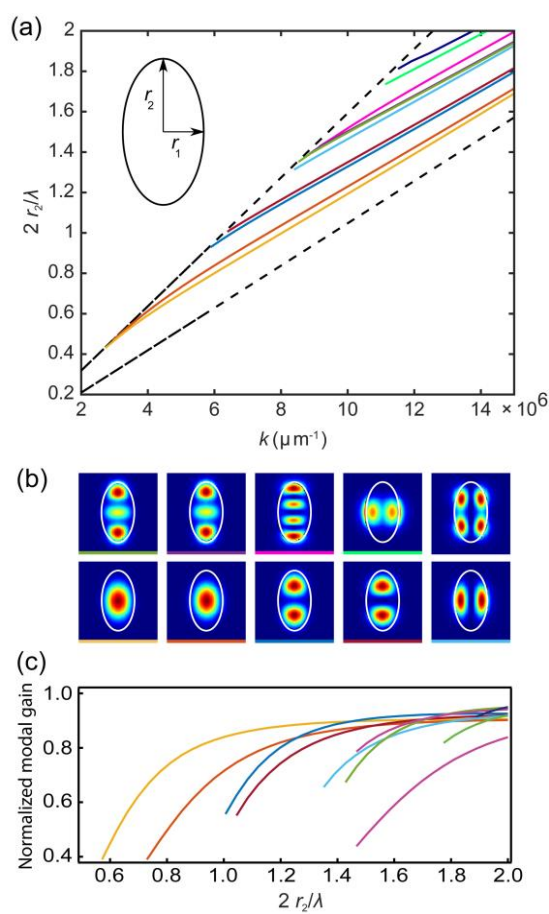

**Figure 6.** (a) Dispersion of guided modes supported by the DCNP/DNA-CTMA nanofibers, *vs.* transverse momentum ($k$), calculated for varying fiber size. Each color line represents a mode which is resonant for large enough fibers, and has a well-defined momentum. The calculations are performed for a constant ellipticity ($r_2 / r_1 = 0.5$), and varying $r_2$ up to one wavelength. The two dashed lines correspond to the light-line in DNA (lower) and in air (upper), which bound the guided modes in the fibers to be in the middle region. Inset: Nanofiber transverse geometry, and relevant dimensions. (b) Transverse intensity profiles of the modes shown in (a). The various modes are identified by the bottom colored bar, with the same colors as in (a), and computed for calculate for $r_1$=0.25 µm and $r_2$ = 0.5 µm. (c) Modal gain, calculated for the same modes and displayed with the same color, and for varying fiber sizes, normalized to the gain of a corresponding homogenous medium.

**Table 1.** Absorption ($\Delta E_{ABS}$) and fluorescence ($\Delta E_{FL}$) energies calculated for DCNP in gas phase and in ACN solution. The experimental data are shown in parentheses (here associated with the most stable isomer). $\Delta\Delta G$ refers to the difference in energy of the *S-cis* isomer compared to the *trans* isomer. The energies are in eV.

|  | $\Delta\Delta G$ | $\Delta E_{ABS}$ | $\Delta E_{FL}$ | Stokes Shift |
|---|---|---|---|---|
| *S-trans* | | | | |
| Vacuum | | 3.20 | 2.81 | 0.39 |
| ACN | | 2.93 (2.70) | 2.54 (2.28) | 0.39 (0.42) |
| *S-cis* | | | | |
| Vacuum | 0.134 | 2.97 | 2.64 | 0.34 |
| ACN | 0.086 | 2.72 | 2.36 | 0.36 |

**Table 2.** CC=CC: torsional dihedral angles in the ground ($S_0$) and excited ($S_1$) state optimized structure of DCNP in the two intercalation modes. $\Delta\Delta G$: Gibbs free energy difference between the two intercalation modes in the ground and excited state. Absorption ($\Delta E_{ABS}$) and fluorescence ($\Delta E_{FL}$) energies in the two intercalation modes. Angles are in degrees, energies in eV.

|  |  | CC=CC | $\Delta\Delta G$ | $\Delta E_{ABS}$ | $\Delta E_{FL}$ |
|---|---|---|---|---|---|
| $S_0$ | Sandwich-like | 178 | 0.000 | 2.93 | |
|  | Rotated | 172 | 0.171 | 2.98 | |
| $S_1$ | Sandwich-like | 176 | 0.000 | | 2.51 |
|  | Rotated | 168 | 0.051 | | 2.55 |

Supporting Information

# Dye stabilization and wavelength tunability in lasing fibers based on DNA

*Luana Persano[1], Adam Szukalski[2], Michele Gaio[3], Maria Moffa[1], Giacomo Salvadori[4], Lech Sznitko[2], Andrea Camposeo[1], Jaroslaw Mysliwiec[2], Riccardo Sapienza[3], Benedetta Mennucci[4], Dario Pisignano [1,5]*

[1]NEST, Istituto Nanoscienze-CNR and Scuola Normale Superiore, Piazza S. Silvestro 12, I-56127 Pisa, Italy.

[2]Faculty of Chemistry, Wroclaw University of Science and Technology, Wybrzeze Wyspianskiego 27, 50-370 Wroclaw, Poland.

[3]The Blackett Laboratory, Department of Physics, Imperial College London, London SW7 2AZ, UK.

[4]Department of Chemistry, University of Pisa, Via G. Moruzzi 13, I-56124 Pisa, Italy.

[5]Dipartimento di Fisica, Università di Pisa, Largo B. Pontecorvo 3, I-56127 Pisa, Italy.

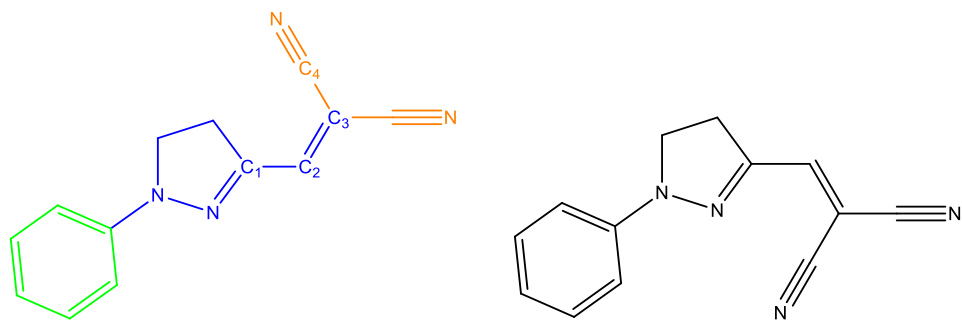

**Figure S1.** Molecular structure of [3-(1,1-dicyanoethenyl)-1-phenyl-4,5-dihydro-1H-pyrazole] (DCNP). Green: electron donor sub-unit, red: electron acceptor sub-unit. Blue: network of conjugated bonds that connect the two parts. The *S-trans* isomer and the *S-cis* isomers are shown in the left and in the right part of the figure, respectively.

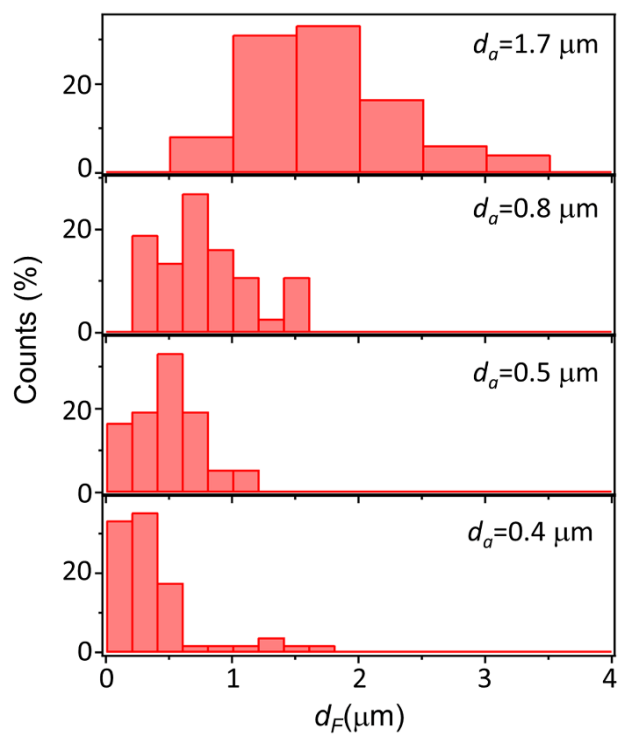

**Figure S2.** Distribution of the transversal size, $d_F$ (~$r_2$, long cross-sectional axis) of DCNP/DNA-CTMA fibers. From top to bottom, the DNA-CTMA concentration in the solution used for electrospinning is 4%, 3.5%, 3%, 2% w/w, respectively. $d_a$: average value of $d_F$ measured for the various distributions.

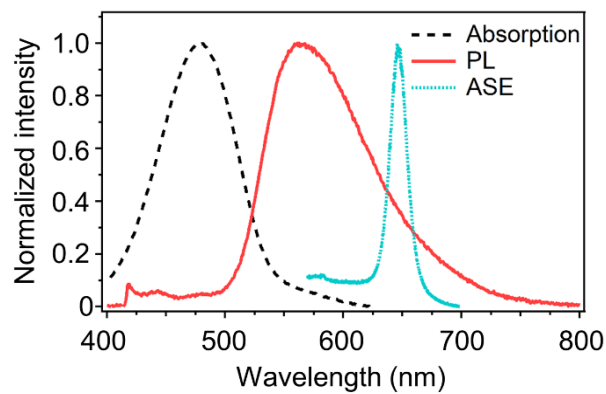

**Figure S3.** Absorption (black dashed line), photoluminescence (PL, red continuous line) and amplified spontaneous emission (ASE, blue dotted line) spectra of DCNP in DNA-CTMA.

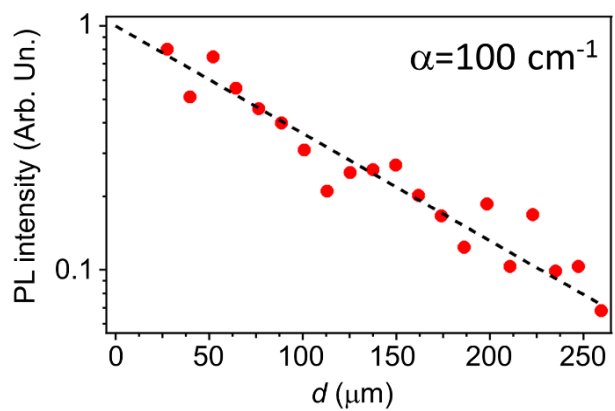

**Figure S4.** Decay of the light intensity (circles) guided along a DCNP/DNA-CTMA nanofiber, as a function of distance, $d$, from the excitation spot. The fiber is excited by a tightly focused laser beam. The dashed line is a fit to the data by an exponential function, $I_{PL}=I_0\times\exp(-\alpha d)$. The obtained loss coefficient, $\alpha$, is about 100 cm$^{-1}$.

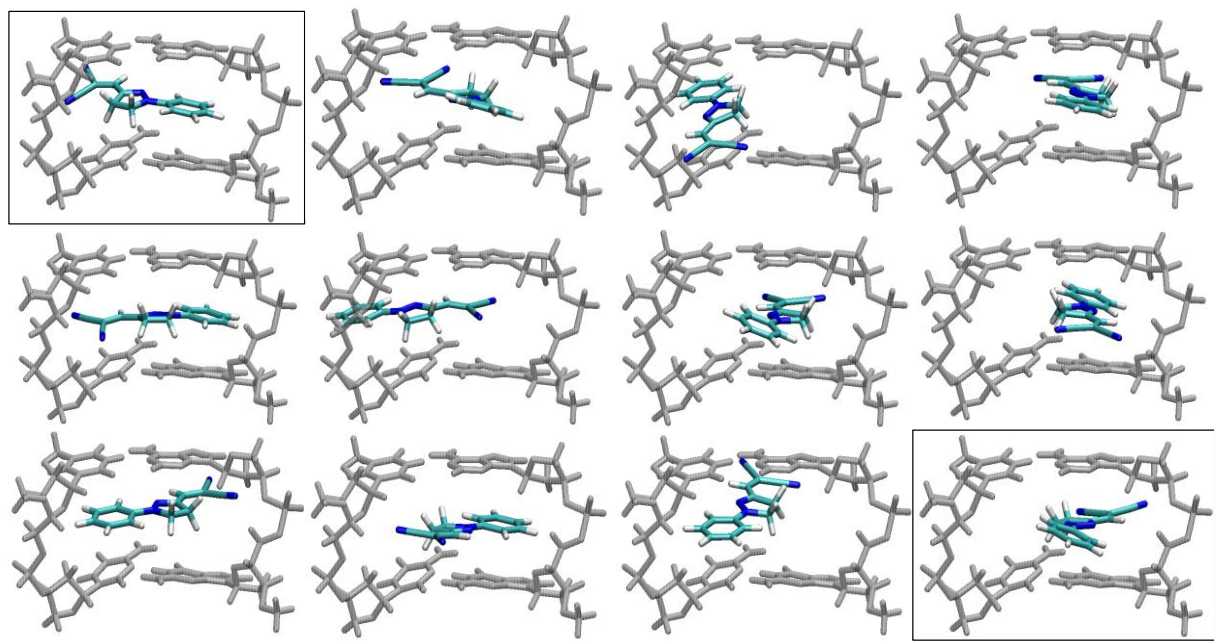

**Figure S5**. Representation of the 12 structures obtained by an initial screening, performed at semiempirical level (PM6), of sandwich-like and rotated intercalation modes. In grey we show the DNA model, and in color the DCNP. The two structures in the box are the two showing the lowest energy for the sandwich-like and for the rotated configuration, respectively. These two structures were finally re-optimized at DFT level (B3LYP/6-31G(d)) with the addition of empirical dispersion corrections.